\documentclass[acmtog,nonacm]{acmart}
\usepackage{braket}
\usepackage{xcolor}
\usepackage{amsmath}
\usepackage{multirow}

\newcommand{\pcircle}{\gamma}

\newcommand{\pf}[1]{{\color{red} }}
\newcommand{\bs}[1]{\boldsymbol{#1}}

\AtBeginDocument{%
  }

\setcopyright{acmlicensed}
\copyrightyear{2026}
\acmYear{2026}
\acmDOI{XXXXXXX.XXXXXXX}
\acmConference[SIGGRAPH 2026]{SIGGRAPH 2026}{July 19--26,
  2026}{Los Angels, CA}
\acmSubmissionID{1667}

\begin{document}

\title{Rendering Coherent Scattering via Quantum Collision Models}

\author{João S. Ferreira}
\email{joao@mothquantum.com}
\orcid{0000-0003-1054-9518}
\affiliation{%
  \institution{Moth Quantum}
  \city{Arlesheim}
  \state{Basel}
  \country{CH}
}

\author{Spencer S. Topel}
\email{spencer@mothquantum.com}
\orcid{0009-0000-0081-9786}
\affiliation{%
  \institution{Moth Quantum}
  \city{Brooklyn}
  \state{New York}
  \country{USA}
}

\author{Pierre Fromholz}
\email{pierre@mothquantum.com}
\orcid{0000-0001-6822-2337}
\affiliation{%
  \institution{Moth Quantum}
  \city{Arlesheim}
  \state{Basel}
  \country{CH}
}

\author{James R. Wootton}
\email{james@mothquantum.com}
\orcid{0000-0003-1943-5306}
\affiliation{%
  \institution{Moth Quantum}
  \city{Arlesheim}
  \state{Basel}
  \country{CH}
}

\renewcommand{\shortauthors}{Ferreira et al.}

\begin{abstract}
Traditional light rendering techniques treat the optical properties of materials as static, yet this assumption breaks down in cases where these properties dynamically evolve in response to incident illumination. We present a novel shading framework that combines classical ray-tracing with a quantum collision model to explore the effect of coherent light-matter interactions in rendering. By treating incident light and material excitations as quantized modes, we model sub-surface scattering as a sequence of symmetry-constrained unitary collisions. This formulation allows for the incorporation of non-integrable dynamics and \emph{chaotic} optical responses due to multi-layer interference effects. We demonstrate how these collision operators can be pre-computed using near-term quantum computers to generate standard BSDFs, enabling the rendering of new physics-inspired materials with distinct optical signatures.
\end{abstract}


\begin{teaserfigure}
  \includegraphics[width=1\textwidth]{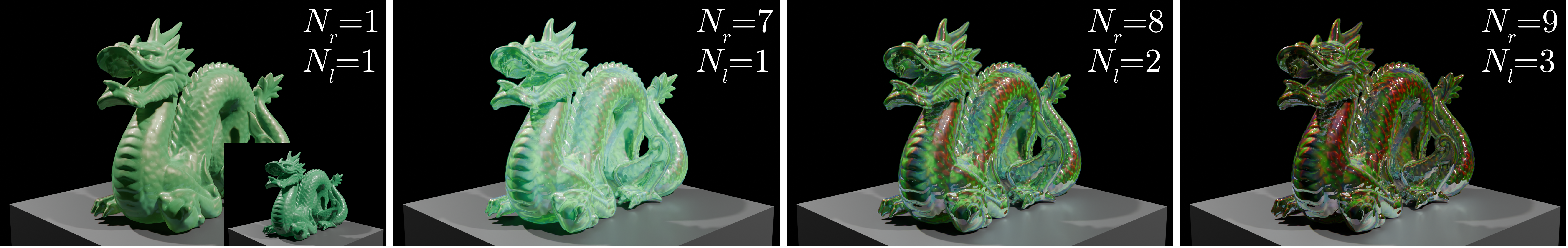}
  \caption{An opaque jade statue (inset) is coated with an offset layer of a novel material node generated using an interacting quantum collision model. From left to right, we increase the number of layers ($N_l$) and rays ($N_r$) creating a richer scene with deeper color contrast due to the presence of quantum interactions.}
  \Description{Stanford dragon rendered with a coat of our QCM material for and increasing number of layer showing more color depth.}
  \label{fig:dragon}
\end{teaserfigure}

\maketitle

\section{Introduction}
Recent advances in nanofabrication have enabled the creation of meta-materials whose optical appearance is governed by coherent propagation, strong light-matter coupling, and geometric properties. Unlike traditional dielectrics, these systems cannot be accurately described by the statistical approximations of geometric optics, as their visual signature depends on the coherent history of the photon traversing an adaptive medium. Current rendering techniques are not equipped to model these effects since absorption is treated as an incoherent energy loss rather than a coherent amplitude, and non-linear interactions between light modes are not supported; both of which are essential to reproduce the optical behavior of strongly coupled quantum materials.

To bridge the gap between macroscopic ray tracing and microscopic wave solvers, we present a rendering framework based on Quantum Collision Models (QCMs). Our approach reinterprets propagating rays and material excitations as quantum particles that interact locally in space and time. By modeling light-matter interaction as a discrete sequence of \emph{elastic} scattering events, we can explicitly track the evolution of the material's internal state (its \textit{memory}) in response to incident illumination. This formulation preserves fundamental symmetries while naturally capturing the complex interference phenomena that arise when the medium adapts to the light passing through it.
Our method is inspired by the physics of multi-layered 2D materials, such as graphite, where optical excitations are confined to individual planes. By explicitly modeling these local interactions, we obtain a flexible framework in which absorption enters as a coherent amplitude on equal footing with reflection and transmission, enabling both traditional interference effects and more general non-linear interaction mechanisms. From a graphics perspective, this allows for the generation of new materials or the augmentation of existing models with physically motivated controls. Furthermore, in regimes involving many layers where classical evaluation becomes computationally demanding, our approach remains viable on current quantum hardware as it only requires the precomputation of optical properties.

This paper makes the following key contributions:
\begin{itemize}
\item A collision-based interaction model that extends ray tracing to include coherent absorption and nonlinear light-matter effects.
\item The derivation of symmetry-constrained unitary operators suitable for simulating the interface of 2D materials.
\item A multilayer interaction framework that produces iridescence through repeated collisions.
\item A practical shader implementation for use in production rendering pipelines. \end{itemize}

\begin{figure*}[ht]
    \centering
    % Placeholder for future figure
    \includegraphics[width=\textwidth]{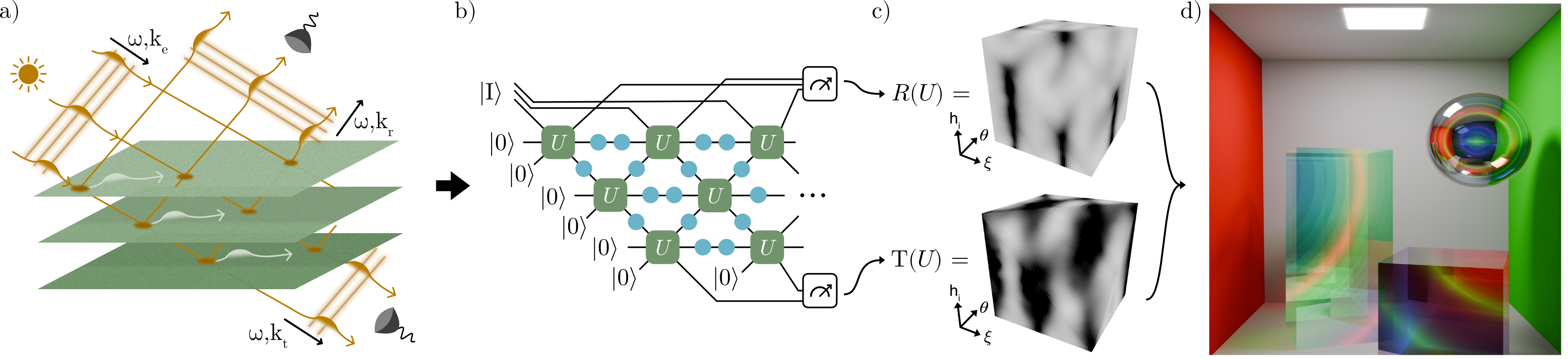}
    \caption{
a) Incident light with frequency $\omega$ and direction $k_e$ interacts with a stack of 2D materials (green). Individual rays are understood as quantum particles that exchange energy with surface excitations, producing a complex reflected and transmitted optical response measured by reflected and transmitted detectors. 
b) The interaction in a) is reinterpreted as a repeated quantum collision model between rays and surface modes, described by a symmetry-preserving unitary ${U}$. 
c) Measurements from b) are used to reconstruct reflectance $R$ and transmittance $T$ as functions of all the microscopic parameters ($h,\theta,...$) and implemented as a multidimensional LUT. 
d) The LUT is used to build a BSDF that yields rich optical responses that strongly vary on the number of layers and rays.}
    \label{fig:collision_schematic}
\end{figure*}
\section{Related Work} \label{sec:related}

Our work lies at the intersection of advanced methods for rendering complex materials and the material science of novel metamaterials. We summarize relevant prior work in these areas here.

\textbf{Light-Matter Interaction in Meta-materials.} Materials such as semiconductors \cite{haug_quantum_2009}, transition metal dichalcogenides \cite{mak_atomically_2010}, and organic crystals \cite{mutter_linear_2007} display strong, non-linear light-matter coupling. In these media, light does not simply reflect or transmit; it interacts with the material, modifying the optical response of both the medium and the field itself. A particularly compelling case arises in materials hosting surface excitons or polaritons which are collective excited modes confined to the interface between two media \cite{paik_excitons_2024,byrnes_excitonpolariton_2014,Roger2015-yw}. Accounting for these modes in the optical response is possible by modifying the Fresnel equations through the introduction of a surface conductivity tensor \cite{oliva-leyva_unveiling_2019}. This conductivity can be derived through various theoretical and empirical methods, including microscopic electron theory \cite{haug_quantum_2009} and equation-of-motion approaches \cite{chaves_excitonic_2017}. We adopt this formulation as the backbone of our algorithm, leveraging it to physically model complex optical behaviors.

\textbf{Rendering Iridescence and Thin Films.}  Iridescent phenomena have been extensively studied in computer graphics, particularly regarding thin films, soap bubbles, and biological structures. Early work by Dias \cite{dias_ray_1991} and Smits \cite{smits_newtons_1992} laid the foundation for simulating interference colors. These approaches have since been extended for efficiency and realism, utilizing closed-form expressions \cite{sun_interference_2008}, fast RGB spectral approximations \cite{belcour_practical_2017}, and integration with microfacet theory or irregular surfaces \cite{belcour_practical_2017,kneiphof_realtime_2019}. Traditional iridescence is fundamentally a many-photon phenomenon, most naturally described within the coherent-state formalism. To the best of our knowledge, there has been no prior computer graphics study of iridescence in the regime of very few photons, which is precisely the focus of this work.

\textbf{Complex Structures and Absorption.} Beyond simple thin films, complex structures such as patterning \cite{gartstein_guiding_2018} or stacking \cite{blackstone_van_2021}, Bragg mirrors \cite{fourneau_interactive_2024}, and pearlescent materials \cite{guillen_general_2020} have been simulated to reproduce interference effects. However, these methods typically focus on wave-optics interference while simplifying or ignoring the effects of strong light-matter interaction and absorption. Alternatively, the presence of stacking or patterning can be directly accounted for in the calculation of surface conductivity \cite{gartstein_guiding_2018}. This approach  can partially capture strong light-matter interaction but still does not incorporate effects originating from absorption. Our approach differs by explicitly exploring how absorption influences the optical response in regimes of strong light-matter coupling, opening new avenues for physical synthesis.

\section{Overview}
\label{sec:overview}

The setup shown in Fig.~\ref{fig:collision_schematic}-a) serves as the inspiration for our reformulation of light-matter interactions. We consider multiple stacked layers of 2-dimensional materials uniformly separated. An incident electromagnetic wave impinges on the material with the wave vector $\bs{k}_e$, creating surface excitations while simultaneously generating reflected and transmitted waves.

Conceptually, the resulting electromagnetic (EM) waves bounce back and forth between the interfaces, tracing optical paths determined by classical Fresnel optics. We assume the propagating rays interact locally with an interface only at the specific instant the optical path intersects the surface. Crucially, the state of the interface evolves between collisions. We assume surface excitations are long-lived, meaning the medium retains a \textit{memory} of previous interactions. This allows us to account for the combined effects of absorption and interference between different optical paths, enabling complex modulation of the outgoing reflected and transmitted fields.

To account for the superposition of all possible optical paths and surface configurations, particularly when the medium's state is entangled with the EM field, a quantum mechanical treatment is natural. In the following sections, we demonstrate how the physical picture in Fig.~\ref{fig:collision_schematic}-a) can be mapped to the equivalent quantum circuit shown in b) and used to render new synthesized materials as in d).

\begin{figure}[t]
    \centering
    \includegraphics[width=0.8\columnwidth]{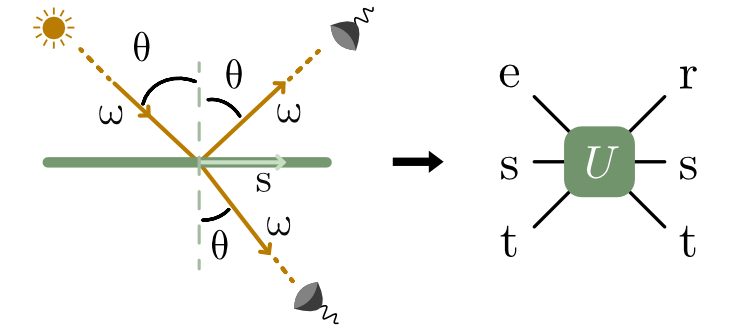}
    \caption{ A single emitted ray, $e$, with a frequency $\omega$ interacts with the surface to produce reflected ($r$), transmitted ($t$) and surface ($s$) modes. The transformation can be re-written as a quantum collision encoded though a unitary $ {U}$ that respects strong symmetry constraints.}
    \label{fig:LM_collision_schematic}
\end{figure}

\section{Light-Matter Interaction of a Single Collision}

We now introduce the formalism and working hypothesis used in our QCM, and explore in depth the case of a single collision characterized by a single matrix $U$. Our end goal is the composition of several of these $U$, explored in Sec.~\ref{sec:multi_collision}.

\subsection{Background: Second Quantization Notation}
\label{sec:background}
Before detailing our method, we briefly introduce the formalism of second quantization~\cite{Mandel2013-tq}, which provides a unified language for describing discrete excitations in both EM fields and matter. 
In this framework, the state of a system supporting $n_{\pcircle}$ excitations of a specific mode $\pcircle$ is denoted by the Fock state $\ket{n_{\pcircle}}$. 

For the electromagnetic field, these excitations correspond to photons and $\pcircle$ denote a certain frequency ($\omega$), momentum and polarization of those photons. For materials, it will correspond to localized modes such as polaritons or excitons. If two distinct excitations, $\pcircle$ and $\pcircle'$, are not correlated, the combined Fock space is the tensor product of each $\ket{n_{\pcircle}}\otimes\ket{m_{\gamma'}}=\ket{n_{\pcircle}m_{\pcircle'}}$. 

Transitions between states are generated by the creation and annihilation operators, $b^\dagger$ and $b$, which add or remove a single excitation,
\begin{align}
b^{\dagger}_{\pcircle}\ket{ n_{\pcircle}} &= \sqrt{n_{\pcircle}+1}\ket{ (n+1)_{\pcircle}} \\
b_{\pcircle}\ket{ n_{\pcircle}} &= \sqrt{n_\pcircle}\ket{ (n-1)_{\pcircle}}\;.
\end{align}
 The number of excitations in each mode is given by the expectation value $\bra{S}b^{\dagger}_{\pcircle}b_{\pcircle}\ket{S}$ (or $\braket{b^{\dagger}_{\pcircle}b_{\pcircle}}_S$ for short) of the number operator $b^{\dagger}_{\pcircle}b_{\pcircle}$ with respect to the target state $\ket{S}$.

This representation naturally supports states that are in superposition such as
\begin{equation}
\alpha \lvert 1_{\pcircle } \rangle + \beta \lvert 0_{\pcircle } \rangle\;,
\end{equation}
which are responsible for the interference effects central to the phenomena explored in this work.

The state of a quantum system evolves as it propagates and interacts with materials. Specifically, an initial state $\ket{I}$ evolves to $e^{-iH\delta t}\ket{I}$ after a short time $\delta t$, where the unitary operator $e^{-iH\delta t}$ is called the evolution operator and $H$ is the Hermitian operator known as the Hamiltonian, which governs the system's dynamics. Any time evolution, including the collision of two systems can be described as sequence of unitaries, e.g. a photon hitting a material.

\subsection{Single Collision of a Single Ray}\label{sec:single}
We start with a single incoming ray made of a single photon with frequency $\omega$ and incident direction $\theta$, relative to the normal of a free-floating conductive 2D material, see Fig.~\ref{fig:LM_collision_schematic}. In future work, we will extend the formalism to account for different medium permittivities but for now let us assume vacuum above and below the surface. 

In second quantization, the state of the system prior to the ray $e$ hitting the surface is given by\footnote{$r$ and $e$ designate the same incident/emitted mode which becomes the reflected mode after the collision} $\ket{I} = \ket{1_{e} 0_t 0_s}$, where the EM field below the surface $t$ (transmitted) and the collective modes of the surface $s$ start out empty.

After the collision, the weight of the incoming mode is split in the three outgoing modes according to the complex scattering amplitudes $r, t, a$
\begin{align}
    \ket{F} = r\ket{1_r0_s0_t} +  a\ket{0_r1_s0_t}+t\ket{0_r0_s1_t}
\end{align}
that must satisfy the energy and momentum conservation 
\begin{align}\label{eq:unitarity}
    R+T+A=1
\end{align}
where $A$ is the absorption and the reflectance $R$ and transmittance $T$ follow directly from the mode's occupations in the final state,
\begin{align}
R= \frac{\braket{b^{\dagger}_{r} b_{r}}_F}{\braket{b^{\dagger}_{r} b_{r}}_I},\; T=\frac{\braket{b^{\dagger}_{t} b_{t}}_F}{\braket{b^{\dagger}_{r} b_{r}}_I}\;
\label{eq:LM_RT}
\end{align}
as in normal optics, see App.~\ref{app:2nd}. We implicitly assumed that the surface mode carries the same energy $\omega$ as the rays. Extensions to account for different frequencies are possible provided we change Eq.~\eqref{eq:LM_RT} accordingly. 

After the collision, all the modes continue to propagate for a time $\tau$ after the collision, with the state picking up a phase build up $e^{-i \omega \tau (b^\dagger_{r}b_{r}+b^\dagger_{t}b_{t}+b^\dagger_{s}b_{s})}\ket{F}$. The angle of propagation remains unchanged after the collision since the materials above and below the 2D surface are the same.

\subsection{The Scattering Matrix}\label{sec:Smatrix}

More generally, any scattering event, whether involving a single ray or multiple incident rays, is encoded as a unitary operator $U$ such that $\ket{F} = U\ket{I}$, where $\ket{I}$ and $\ket{F}$ are the pre- and post-scattering states.

The operator $U$, must be a valid scattering matrix and thus satisfy some constraints. First, any expectation value derived from the post-collision state must be invariant under the simultaneous swap of the reflection and transmission indices ($S_{r\leftrightarrow t}$), i.e.
\begin{align}
 {U}S_{r\leftrightarrow t}&=S_{r\leftrightarrow t}{U}\;.
\label{eq:sym_flip}
\end{align}
In other words, the same $ {U}$ should describe both an incidence from above and below.

%If we wish to model a material with known $R, T, A$, it constrains the scattering amplitudes $r, t, a$ thus fixing $U$. Conversely, we can explore the optical response of any valid $U$ satisfying our symmetry constraints, see Sec.~\ref{sec:symmetry}.

Second, the collision must conserve the total energy of the system, %e.g. Eq.~\eqref{eq:unitarity}
which in our formulation corresponds to the conservation of the number of excitations. This implies that $U$ is block-diagonal in the excitation-number basis:
\begin{equation}
    U = \bigoplus_{n\geq 0} U_n\;,
    \label{eq:block_diagonal}
\end{equation}
where $U_n$ acts on the $n$-excitation sector of the Fock space. The vacuum sector ($n=0$) is trivial, the state $\ket{0_r 0_s 0_t}$ must always remain unaltered so $U_0=1$. The single-excitations sector $U_1$ encodes the scattering of the states $\{\ket{1_r 0_s0_t}$, $\ket{0_r 1_s0_t}$, $\ket{0_r 0_s 1_t}\}$. In that basis, the collision matrix reads:

\begin{equation}
    U_1 = \begin{pmatrix} r & a & t \\ a & c & a \\ t & a & r \end{pmatrix}\;,
    \label{eq:U1_matrix}
\end{equation}
where $c=-a(\bar{r}+\bar{t})/\bar{a}$. The first column correspond to the collision of Sec.~\ref{sec:single} while the second column corresponds to the process of a material excitation decaying into two photons. 

Terms $U_2$ and above incorporate scattering processes between multiple rays which, in the presence of nonlinearities, cannot be expressed as a function of only $r,t,a$. Since the Fock space is not upper bounded, $U$ has infinite dimensions. 

The most practical way to describe $U$ is using the \emph{collision Hamiltonian}
\begin{equation}
    H^c = i\log U= h_k\sum_k H_k ,
    \label{eq:collision_H}
\end{equation}
which generates $U = e^{-iH^c}$ and captures the full structure of the collision. $H^c$ can be expanded as sum of independent terms, $H_k$, that themselves also satisfy the aforementioned symmetries. Some of these are shown in Tab.~\ref{tab:hamiltonians} and grouped by excitation sector. Each term can be understood as an effective coupling between light and surface; for example, $H_2$ pertains to transmission of a ray without absorption, $H_3$ pertains to absorption/emission of the incident ray by the surface while terms $H_4$ and above represent different processes mediated by the presence of an additional mode.

If our goal is to model a linear material, only $H_{0-3}$ will be non-zero while nonlinearities arise whenever any interaction terms, $H_{\geq4}$, are present.

\begin{table}[]
    \caption{The first terms of the collision Hamiltonian $H^c$ organized by the excitation-number sector.}
    \centering
    \begin{tabular}{|c|c|p{4.5cm}|}
    \hline
    \multirow{4}{*}{$U_1$} & $H_0$ & $b_{a}^{\dagger}b_{a}$ \\
    \cline{2-3}
    & $H_1$ & $b_{r}^{\dagger}b_{r}+ b_{t}^{\dagger}b_{t}$ \\
    \cline{2-3}
    & $H_2$ & $b_{t}^{\dagger}b_{r}+b_{t}b_{r}^{\dagger}$ \\
    \cline{2-3}
    & $H_3$ & $\left(b_{t}^{\dagger}+b_{r}^{\dagger}\right)b_{s} / \sqrt{2}+h.c.$ \\
    \hline \hline
    \multirow{7}{*}{$U_2$}& $H_4$ & $\left(b^{\dagger}_tb_r+b_tb^{\dagger}_r\right)b_{s}^{\dagger}b_s$ \\
    \cline{2-3}
    & $H_5$ & $\left(b^{\dagger}_tb^{\dagger}_rb_r + b^{\dagger}_rb^{\dagger}_tb_t\right)b_s/ \sqrt{2}+h.c.$ \\
    \cline{2-3}
    & $H_6$ & $b_{t}^{\dagger}b_tb_{r}^{\dagger}b_{r}$ \\
    \cline{2-3}
    & $H_7$ & $\left(b_{t}^{\dagger}b_t+b_{r}^{\dagger}b_{r}\right)b_{s}^{\dagger}b_s /2$ \\
    \cline{2-3}
    & $H_8$ & $\left({b_{t}^{\dagger}}^2+{b_{r}^{\dagger}}^2\right)b_{s}^2+ h.c.$ \\
    \cline{2-3}
    & $\dots$ & $\dots$ \\ % Added ellipsis inside U2
    \hline \hline 
    \multirow{2}{*}{$U_3$} & $H_9$ & $b_{t}^{\dagger}b_tb_{r}^{\dagger}b_{r}b_{s}^{\dagger}b_s$ \\
    \cline{2-3}
    & $\dots$ & $\dots$ \\
    \hline
\end{tabular}
    \label{tab:hamiltonians}
\end{table}
\begin{figure*}[ht]
    \centering
    \includegraphics[width=\textwidth]{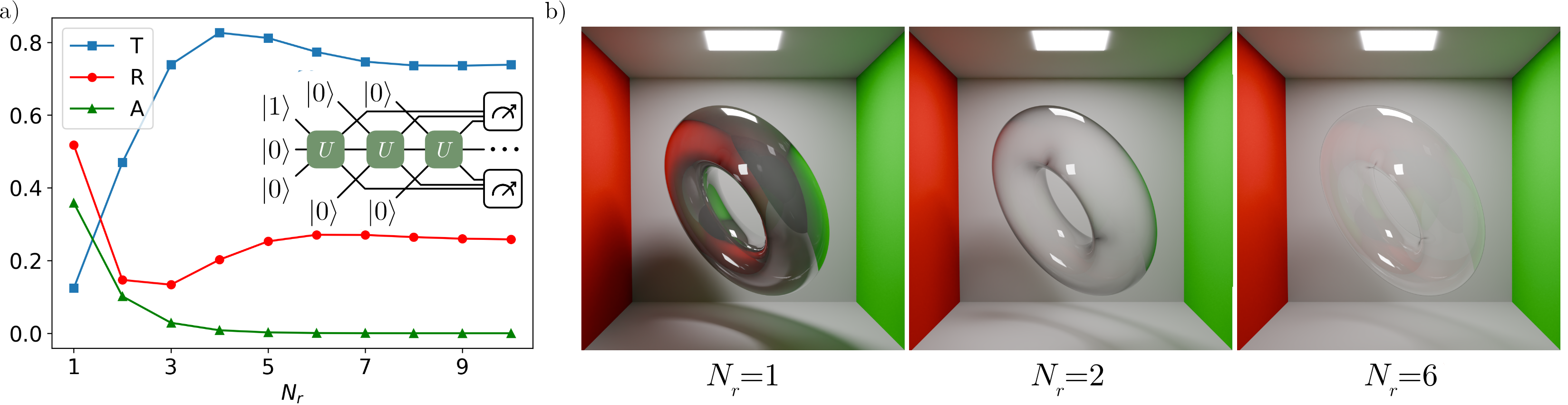}
    \caption{a) For the set-up depicted in the inset, the occupation of the surface mode ($A$) decays exponentially with every added ray while $R,T$ values converge. We consider that re-emission is immediate so there is no phase in between collisions and fix $\theta=3\pi/8$. b) Visual comparison for an increasing number of rays leading to a more translucent material.}
    \label{fig:multi_relaxation}
\end{figure*}
\subsection{Constructing a Valid $U$\label{sec:find_u}}

We illustrate how to construct a valid $U$ for a 2D material. Assuming \emph{p}-polarized light, the generalized Fresnel equations for a 2D conducting surface~\cite{oliva-leyva_unveiling_2019} are given by
\begin{align}
    r=-\frac{\sigma}{2\cos\theta +\sigma},\;t&=1+r
\end{align}
where $\sigma(\omega)$ is the \emph{complex} surface conductivity, which in general depends on the ray's frequency $\omega$. The absorption amplitude is $a=\sqrt{|r|^2+|t|^2}\,e^{i\theta_a}$, up to a model-dependent phase $\theta_a$, which fully fixes the $U_1$ matrix of Eq.~\eqref{eq:U1_matrix}.

With $U_1$ in hand, the coefficients $h_{0\text{--}3}$ follow directly from Eq.~\eqref{eq:collision_H}. For a 2D material with a linear optical response, as in the case of idealized free-standing graphene~\cite{stauber2008optical}, this fully characterizes the collision matrix $U$.

Non-idealized 2D materials, such as graphene nanostructures or samples with impurities, can host strong nonlinear optical regimes. Within our framework, these can be included via higher-excitation interaction terms $H_{\geq 4}$ in the collision Hamiltonian. The precise choice of interactions and their weights would require a microscopic characterization of the target material, which is beyond the scope of this work. 

Instead, we treat the non-linear terms as artistic tools. We make an \emph{artistically motivated} choice composed of two contributions, giving the full unitary
\begin{align}
    U= \exp\left(i\sum_{k=0}^3h_kH_k+ h_i (h_2H_4+h_3H_5 - H_7 - H_9)\right),
\end{align}
with $h_{0\text{--}3}$ fixed from $r,t,a$ and $h_i$ controlling the overall strength of the non-linear components. The first contribution, $h_2H_4+h_3H_5$, doubles the hopping amplitudes between the reflected, transmitted and absorbed modes when additional excitations are present, ensuring that propagation persists when multiple rays collide simultaneously. The second contribution, $-H_7-H_9$, introduces an attractive coupling between modes that further amplifies the effect of multi-ray collisions, making the non-linear signature more pronounced.

The limiting case $U(h_i=0)$ corresponds to free-boson dynamics and can be trivially simulated classically, while $U(h_i\neq0)$ introduces non-trivial many-body effects that give rise to the complex visual phenomena explored in the following sections.

Since our goal is not to render any physically accurate 2D material, we fix $\sigma=1-i$ and $\theta_a=0$ throughout, so that $U$ depends only on two free parameters: the incidence angle $\theta$ and the interaction strength $h_i$. Unless otherwise stated, all results in the remainder of this paper use this parametrization.

\section{Light-Matter Interaction for Multiple Collisions\label{sec:multi_collision}}

Up to this point we have shown that a refraction process can be reformulated as a quantum collision process, where the parameters of the collision unitary can be chosen so to match the reflectance and transmittance of a given material at a specified incident angle.
Besides replicating known results, the formulation enables a natural extension to scenarios involving multiple rays, repeated interactions or relaxation processes whilst accounting for quantum effects such as superposition. 

\subsection{Hardware Considerations}\label{sec:reduction}

As we extend to multi-collision scenarios, we must address a practical concern. The complexity of simulating set-ups where multiple modes interact and become entangled  grows exponentially with the number and dimensionality of the modes involved in the collision, meaning that classical simulation can quickly become intractable. Future quantum computers provide a solution to this problem, allowing us to simulate such collisions with a polynomial number of qubits; however, current hardware is very limited in the number of qubits and logical operations and we must resort to simulating the circuits classically.

In order to keep our procedure computationally tractable, we choose to truncate the Fock space to a maximum of 1 excitation per mode such that
\begin{align}
    b^{\dagger}\ket{1}=b\ket{0}=0\;.
\end{align}
Limiting to a single excitation means that each mode can be described by a single qubit and, consequently, each gate $U$ requires only 3-qubit gates, making the circuits compatible with current quantum hardware which will be the subject of future work.

Discarding states with multiple excitations in the same mode, e.g. $\ket{2_r0_s0_t}$, means that we cannot capture the classical non-linear effects that emerge from having many photons in the same mode, but we can still capture non-trivial quantum effects that emerge when multiple photons from different modes interfere. Notably, the single- and few-photon regime is also the one where quantum effects are most pronounced: superposition and entanglement are strongest precisely when the number of excitations is small. 

This truncation thus suffices to generate interesting creative outcomes and constitutes a simple starting point before extensions accounting for more excitations. 
\begin{figure*}[ht]
    \centering
    \includegraphics[width=0.9\textwidth]{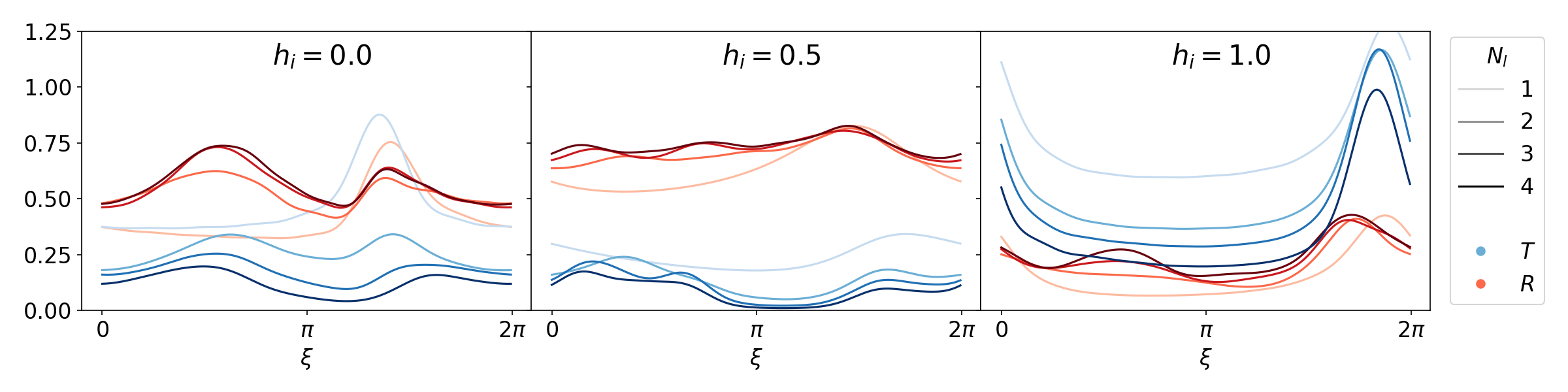}
    \caption{Scaling of $R$ (orange curves) and $T$ (blue curves) for different numbers of layers and rays ($N_r = N_l +6$) and interaction strengths $h_i$ as a function of the phase difference $\xi$. We fixed $\sigma=1-i$ and $\theta=0$.}
    \label{fig:rtcurves}
\end{figure*}
\subsection{Absorption and Re-emission} \label{sec:abs-reem}

As a first multi-collision scenario, we consider a single absorbing layer whose excited mode relaxes through competing channels that may include light re-emission~\cite{cong2018optical}. Within our formalism, this is captured by applying $U$ repeatedly to the same surface mode, each time coupled to a new pair of outgoing modes initialized in the vacuum, generating a sequence of emitted rays that progressively drain the surface excitation towards the ground state, as illustrated in the insert of Fig.~\ref{fig:multi_relaxation}-a). We work in the limit where re-emission is instantaneous such that there is no phase accumulation. The emitted rays are labeled by the extra index $l \in [1,N_r]$, where $N_r$ is the total number of rays considered.

When considering several incident rays instead of just one, the expression for reflectance and transmittance are modified to account for the cross terms between the different outward modes,
\begin{equation}
\braket{b^{\dagger}_{\mathsf{x}}b_{\mathsf{x}}}\rightarrow \sum_{j,k=1}^{N_r}\braket{b^{\dagger}_{\mathsf{x},j}b_{\mathsf{x},l}}\;,
\label{eq:multi_incident_intensity}
\end{equation} where $\mathsf{x}=r,t$ and the indices $j,l$ sums over all the reflected or transmitted  qubits. The expectation value is evaluated on the final state obtained at the same time for all qubits, see  App.~\ref{app:initial_states}\footnote{Since states of the form $\ket{11\dots 1}$ no longer have the highest possible intensity (within each particle-number sector), the coefficients $R$,$T$ and the sum $R+T$  can exceed 1 if the final state is entangled and the initial state is not. }. 

In Fig.~\ref{fig:multi_relaxation}-a), we showcase for a fixed $U(\theta=0.2)$ the exponential decay of the occupation of the excited mode as well as the convergence of $R$ and $T$ when accounting for relaxation processes\footnote{Note that interactions do not play a role here since we are restricted to the single-particle sector. }. The exact $R,T$ curves depend strongly on $\theta,N_r$ so in Fig.~\ref{fig:multi_relaxation}-b) we showcase how this effect can look like in practice for different number of outward rays $N_r$. The case of $N_r=1$ recovers the standard single-collision result with no re-emission, whereas $N_r > 1$ incorporates successive emission events that gradually drain the surface excitation. The overall surface tends to be more translucence and, since we fixed the phase, there is not dependence on $\omega$ ans thus no iridescence pattern.

\subsection{Multiple Layer Set-up}  \label{sec:multi_layer}

Another application is to extend the framework to consider coherent interference effects caused by multiple incident rays on repeated structures. The motivation comes from Van der Waals materials, which consist of atomically thin sheets held together by weak electrostatic forces. The most well-known example is graphite, composed of layers of graphene. These materials can function as optical cavities~\cite{zhang2021interface}, which inspired us to study the configuration of Fig.~\ref{fig:collision_schematic} composed of $N_r$ distinct rays impinging on a stack of $N_l$ uniformly spaced 2D material layers separated by a distance $d$. As rays propagate between collisions, they accumulate a phase $-\omega d/\cos\theta$, while the surface mode acquires twice this amount. Uniform spacing maximizes the coherent interference between light rays and best showcases the role of quantum interactions in the optical response.

\subsubsection{Thin-film }
To build intuition, we first consider the simplest non-trivial case: a single photon incident on two layers, $N_l=2$, described by the input state $\lvert 100\cdots\rangle$. With only one excitation present, interactions play no role and the dynamics are purely linear. Classical thin-film interference then predicts a reflectance obtained by summing over all optical paths,
\begin{align}
    R_\text{2L} = \left|r + t^2 r e^{i\xi} + t^2 r^3 e^{2i\xi} + \cdots + a^2e^{i\xi} + tra^2 e^{i2\xi}+ \cdots \right|^2, \label{eq:2L}
\end{align}
where $\xi=-2\omega d\cos\theta$ is the optical path's phase difference, the first group of terms is the standard Airy reflectance formula, and the remaining terms account for paths involving absorption. In Appendix~\ref{app:thin_film} we show that our circuit construction of $U$ recovers this formula exactly for a single-photon input.
\begin{figure}
    \centering
    \includegraphics[width=\columnwidth]{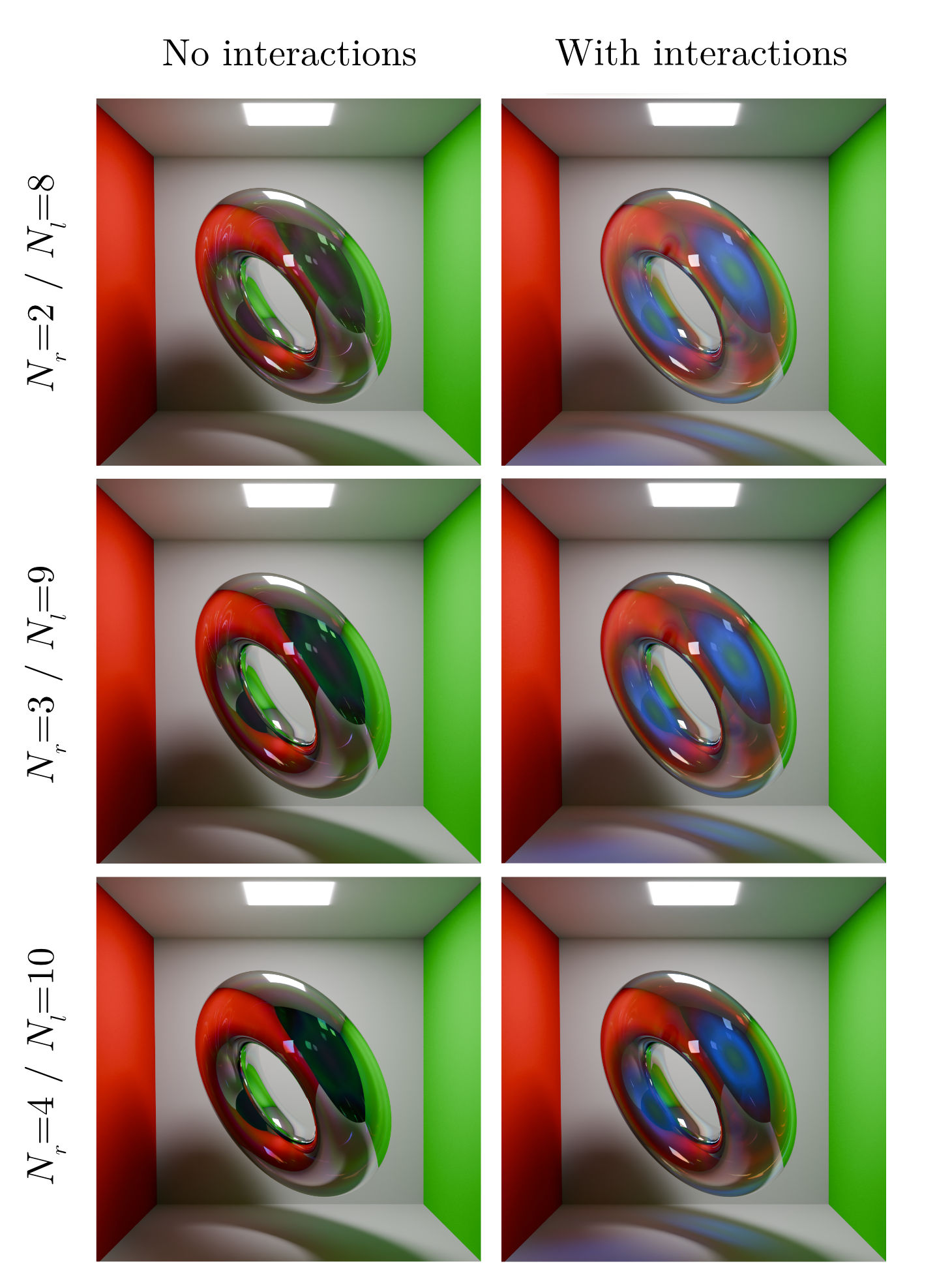}
    \caption{Comparison of the visual outcome for no interactions ($h_i=0$) vs with strong interactions ($h_i=1$) for different number of layers and rays. We fixed $d=250$nm. The presence of non-linear scattering creates a richer iridescence pattern with more colors and deeper contrast.}
    \label{fig:graphite_scaling}
\end{figure}

\subsubsection{Many-Photon Regime}
To probe non-linear effects, one might naturally consider an $N$-photon Fock state $\lvert N\,0\cdots\rangle$ in a single mode. Such states, however, lie outside our truncation of Sec.~\ref{sec:reduction}. The natural analogue within our constraints is instead the family of states $\lvert 11\cdots 1\rangle$, with one excitation distributed across $N_r$ distinct incoming modes. These correlated, multi-photon states are not typical in classical illumination but are experimentally realizable in quantum optics laboratories. 

The full setup is illustrated in Fig.~\ref{fig:collision_schematic}: each of the $N_l$ layers hosts a surface mode initialized to $\ket{0}$, and all modes not directly driven by the incident rays are likewise initialized to vacuum. The optical response is controlled by three parameters: the incidence angle $\theta$, the interlayer distance $d$ (which sets the phase $\xi = -2\omega d\cos\theta$), and the interaction strength $h_i$. Reflectance and transmittance are computed via Eq.~\eqref{eq:multi_incident_intensity} and converge in the $N_r\gg N_l$ limit; unless specified, we use $N_r=N_l+6$.

The visual signature of this configuration strongly depends on the many-body transport properties of light through the layers (i.e. the choice of $U$). For a generic non-linear term in the collision Hamiltonian, one expects quantum-chaotic behavior~\cite{Santos2010-yb}: interactions induce back-scattering and phase mixing between optical paths, which should increase reflectivity as the number of layers grows.

Our specific choice of $U$, however, is far from generic, and Fig.~\ref{fig:rtcurves} reveals how its structured nature leads to richer behavior. At small interactions ($h_i=0.5$), reflectivity increases with $N_l$ as expected from generic backscattering. At stronger interactions ($h_i=1$), however, the structured nature of our choice explicitly favors transmission, leading to a substantial increase in transmittance; for specific values of $\xi$, sharp transmittance peaks emerge that narrow with $N_l$, which may indicate a transition between chaotic and non-chaotic transport regimes, though further study would be needed to confirm this. These interaction-dependent variations translate directly into a richer visual appearance, as shown in Fig.~\ref{fig:graphite_scaling}, where we compare the rendered outcome for no interactions ($h_i=0$) against strong interactions across different $N_r$ and $N_l$ for a fixed $d=250\,\text{nm}$. The scene-dependent interactions not only introduce new colors absent from the original scene, akin to thin-film iridescence, but also yield increasingly rich color variations and deeper contrast as the number of layers grows.

\begin{figure*}[t]
    \centering
    \includegraphics[width=0.95\textwidth]{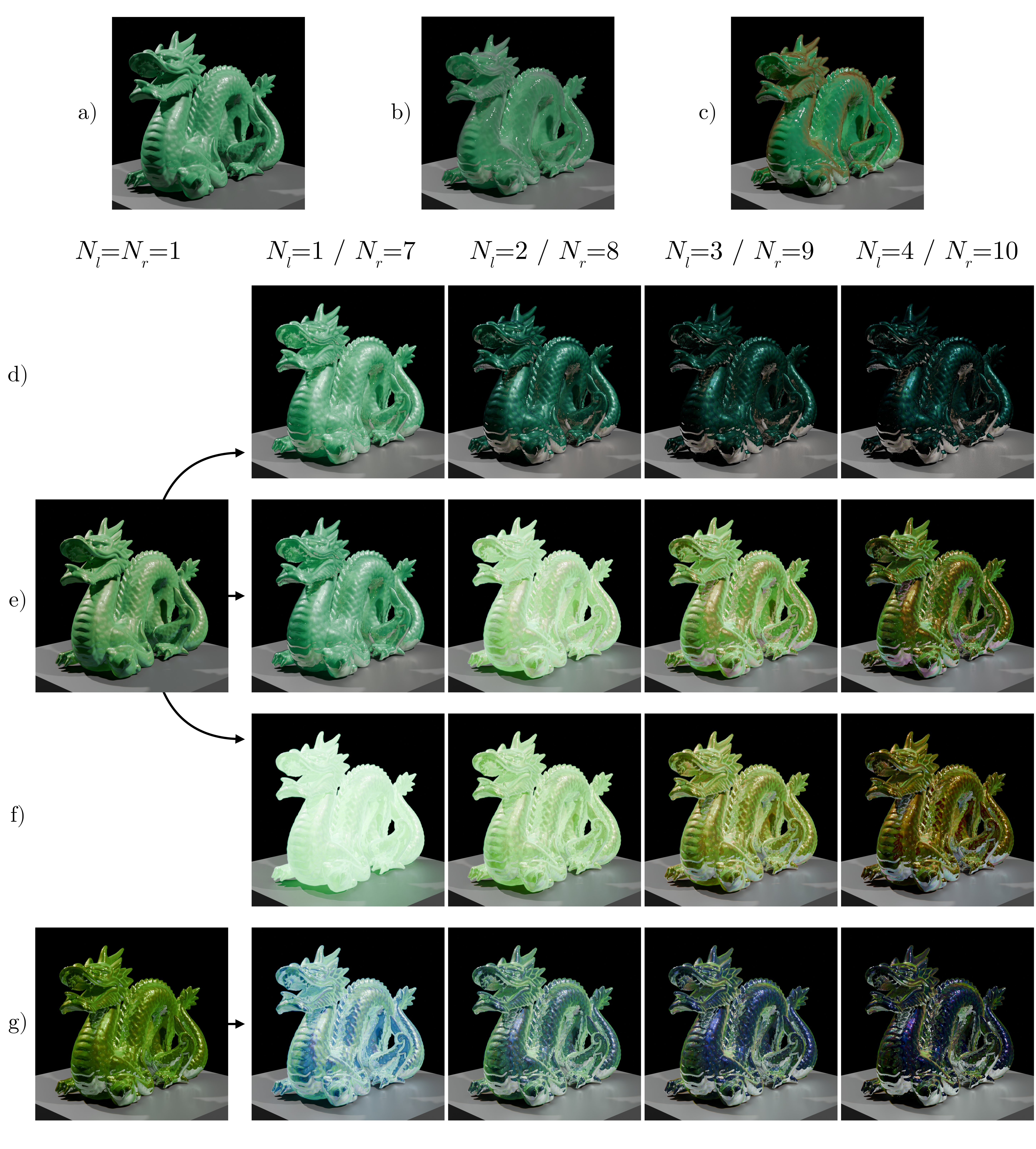}
    \caption{Visual comparison of the effect of different iridescence methods. A baseline jade statue of the Stanford dragon (a) is coated by a material surface of the same shape, offset by 5\%, where we apply the new effects. (b) Standard thin-film iridescence~\cite{belcour_practical_2017} for a thickness of 500 nm and index of refraction 1.5. (c) Hand-generated iridescence effect using a green–red–green color ramp linked to the surface facing. (d)–(f) Effect of the QCM for different numbers of layers $N_l$ and rays $N_r$, generated using the same shader with interaction strength $h_i=1$ but varying interlayer distance. (d) $d=250\,\text{nm}$. (e) $d=350\,\text{nm}$. (f) $d=600\,\text{nm}$. (g) Custom choice of interactions and a frequency-dependent conductivity $\sigma(\omega)$.} 
    \label{fig:dragons}
\end{figure*}

The choice of which effective non-linear terms to use $h_{\mathsf{4-9}}$, provide a flexible, physically grounded control mechanism that artists can tune to create a wide range of visual effects. In Figs.~\ref{fig:dragons}, we demonstrate how different parameters create distinct and visually appealing iridescence patterns and, in Fig.~\ref{fig:dragon}, we show an example of a different non-linear term that depends on $\theta$.

%Exploring the regime of a large number of layers is essential for the visual identity of our approach. However, due to the entanglement in the circuit, the classical simulation cost scales exponentially with $2(N_r + N_l)$ (while it is linear in $N_r.N_l$ for a quantum device, notwithstanding error correction), making such simulations quickly intractable on conventional hardware. Quantum processors are therefore required to access this regime. Implementing our circuit on a quantum computer will be the subject of future work; we note, however, that it only requires 3-qubit gates, making it compatible with current quantum hardware.

\section{Implementation in a Rendering Pipeline}

To integrate this model into a computer graphics workflow, we implemented a custom OSL shader to use with Blender Cycles. The workflow proceeds as follows: the number of layers $N_l$, the number of rays $N_r$, and the conductivity of each layer are chosen upfront, and $R$ and $T$ are then pre-computed for all remaining free parameters (incidence angle, phase, interaction strength, etc.) on a uniform grid of 60 points per parameter. Notably, the phase $\xi$ does not need to be sampled independently, owing to a numerical trick in the circuit evaluation (see App.~\ref{app:initial_states}). As a concrete example, the largest configuration considered here ($N_l=4$, $N_r=10$) requires 24 qubits and takes approximately 12\,s per parameter on an Apple M3 processor using an exact statevector simulator. We note that the circuit structure is well-suited to execution on current quantum hardware and that will be the focus of a future work. The resulting multi-dimensional $R$ and $T$ arrays are baked directly into the shader file as flattened arrays, with linear interpolation used at runtime (as an alternative, the arrays could equally be stored as a separate LUT imported by the shader). Inside the shader, $R$ and $T$ are evaluated per RGB channel: the $R$ values drive a perfectly specular BRDF (\textit{Glossy} node in Blender) while the $T$ values feed a transparent BTDF (\textit{Transparent} node).

\section{Conclusion}

We have introduced a rendering framework that reinterprets light-matter interactions as a sequence of symmetry-constrained quantum collisions. Though this approach can be expanded to account for a broad class of quantum phenomena, we focused on a detailed analysis of the regime in which it displays novel results not seen in standard techniques. It successfully bridges the gap between macroscopic ray tracing and microscopic many-body physics, offering a new method for synthesizing materials with complex optical signatures. By encapsulating these dynamics within a modular shader, we have demonstrated that quantum-mechanical effects, such as coherent backscattering and chaotic interference can be effectively integrated into production grade rendering pipelines.

Several promising directions remain for extending this framework. While we focused on a specific choice of non-linear interactions, this represents only one point in a vast parameter space. Different scattering matrices will yield qualitatively different optical responses and scaling behaviors with respect to the number of layers. Systematically exploring these distinct scaling regimes is the rendering equivalent of characterizing phases of matter, offering artists a rich, physically grounded aesthetic palette where unique visual signatures emerge from fundamental interaction rules. Furthermore, while our current setup assumes a homogeneous background, heterogeneity and disorder can be naturally incorporated by varying the collision operator $U$ in space and time.

Linking our results more formally to classical phenomena, such as the iridescence of Bragg mirrors, requires extending the framework to allow layers with distinct permittivities. A complementary direction is to explore illumination states that go beyond the correlated single-photon inputs considered here, in particular states with multiple photons in the same mode, such as $\lvert N\,0\cdots\rangle$. Accessing this regime requires expanding the Fock space beyond the single-occupancy subspace, which in turn enables the system to be driven by non-classical light such as squeezed states — potentially unlocking visual effects that depend explicitly on the quantum statistics of the source rather than just on material geometry.

However, expanding the Fock space to account for additional modes, rays, or layers presents a fundamental challenge, as the required classical resources grow exponentially with the system size. In this work, our reliance on exact statevector simulation restricted us to a maximum of four layers while advanced classical techniques such as tensor networks could extend this range, they inherently trade accuracy for scalability. A more promising avenue is to execute these circuits directly on quantum hardware. The collision circuits derived here are particularly well-suited to the constraints of current devices in terms of qubit count and gate depth. Furthermore, because our approach decouples simulation from rendering, the slower execution speeds of current quantum processors are not a bottleneck. We can leverage quantum hardware in an offline pre-computation stage to generate high-precision LUTs, which are then re-used by standard real-time shaders.

Finally, this work opens the possibility of exploring quantum mechanical effects in alternative rendering paradigms. Techniques such as ray-marching or Gaussian splatting could similarly benefit from these collision-based models, extending the reach of quantum-inspired visuals to volumetric and point-based domains.

We envision this framework not as a means to perform atomic-level simulations of specific materials, but rather as a vehicle to explore novel, physically motivated quantum effects through the capabilities of quantum computing hardware. It is intended not as a replacement for classical optics, but as a new creative instrument. We invite artists and developers to use these tools to design materials that defy classical intuition, exploring a visual space where the rules of quantum mechanics govern the aesthetic experience.

\begin{acks}
We are grateful to Konstantinos Meichanetzidis for many insightful and ongoing discussions.
\end{acks}

\appendix
\onecolumn
\section{Second Quantization of the Electromagnetic field}\label{app:2nd}

In this section, we briefly recap the second quantization of the electromagnetic field in terms of bosonic creation and annihilation operators, a complete treatment can be found in Ref.~\cite{Mandel2013-tq}. The electric and magnetic fields can be expanded in plane wave modes labeled by momentum $\bs{k}$ and polarization index $\mu$. The corresponding operators $b_{\mu,\bs{k}}$ and $b_{\mu,\bs{k}}^{\dagger}$ create and annihilate photons in these modes and generate the full photonic Fock space.

The electric field operator in a medium is thus proportional to
\begin{align}
\boldsymbol{E}(\bs{r},t)&=\boldsymbol{E}^{(+)}(\bs{r},t)+\boldsymbol{E}^{(-)}(\bs{r},t)\\&\propto i\sum_{\bs{k},\mu}\sqrt{\omega_{k}}\left(\boldsymbol{v}_{\mu,\bs{k}}b_{\mu,\bs{k}}(t)e^{i\bs{k}.\bs{r}}-\bar{\boldsymbol{v}}_{\mu,\bs{k}}b_{\mu,\bs{k}}^{\dagger}(t)e^{-i\bs{k}.\bs{r}}\right)
\label{eq:EM_Efield}
\end{align}
where $\boldsymbol{v}_{\mu,\bs{k}}$ denotes the polarization vectors, $\bs{k}$ the momentum, and $\omega_{k}$ is the dispersion relation in that medium. 

The detection probability for a single-photon detector located at position $\bs{r}_0$ at a time $t_{0}$ along the direction $\mathbf{n}$ is proportional to~\cite{glauber}
\begin{align}
\braket{\boldsymbol{E}^{(-)}(\bs{r}_0,t_0) \cdot \boldsymbol{E}^{(+)}(\bs{r}_0,t_0)}
\propto     \sum_{k,k',\mu,\mu'}\sqrt{\omega_k\omega_k'} \braket{ b_{\mu,\bs{k}}^{\dagger}b_{\mu',\bs{k}'}} \bar{\bs v}_{\mu,\bs{k}}.\bs{v}_{\mu',\bs{k}'}e^{i(k-k')\mathbf{n}.{\bs r}_0}e^{-i(\omega_k-\omega_{k'})t_0} \nonumber
\label{eq:EM_commutation}
\end{align}
where we switched from the Heisenberg picture to the Schr\"odinger picture $b_{\mu,\bs k}(t)\rightarrow b_{\mu,\bs k}e^{i\omega_k t}$ so the expectation value is evaluated on the initial state. The mode operators satisfy the bosonic commutation relations
\begin{align}
\left[b_{\mu,\bs{k}},b_{\mu',\bs{k}'}^{\dagger}\right]=\delta_{\mu\mu'}\delta_{\bs{k}\bs{k}'}
\end{align}

The expression \eqref{eq:EM_commutation} simplifies tremendously if we consider the time average (long exposition time) where
\begin{align}
\int dt e^{-i(\omega_k-\omega_{k'})t}\propto \delta_{k,k'}.
\end{align}
Since $\boldsymbol{v}_{ \mu,\bs k}.\bar{\boldsymbol{v}}_{\mu',\bs k}=\delta_{\mu\mu'}$ the contribution of each mode decouples and we are left with
\begin{align}
\int \mathrm{d}t\braket{\boldsymbol{E}^{(-)}(\bs{r}_0,t) \cdot \boldsymbol{E}^{(+)}(\bs{r}_0,t)}
&\propto \sum_{\mu, k} \omega_{k} \braket{ b_{\mu,\bs{k}}^{\dagger}b_{\mu,\bs{k}}},
\label{eq:EM_simple}
\end{align}
which is understood as the average intensity of the EM field at a given point in space.

The average irradiance on a surface $S$ is thus obtained as 
\begin{align}
I_S \propto \cos(\theta_S)\sum_{\mu,\bs{k}}\omega_{k}  \braket{b_{\mu,\bs{k}}^{\dagger}b_{\mu,\bs{k}}}
\label{eq:EM_intensity_final}
\end{align}
where $\theta_S$ is the angle between the propagation direction and the normal of the surface. The proportionality factor is irrelevant, as we will only consider ratios of irradiance to compute the reflectance and transmittance. 
In this work, we only consider a single polarization but the extension to non-polarized light is straightforward since each contribution can be treated independently.

While the expansion above uses plane waves as a basis, the formalism extends to any complete set of field eigenmodes, $\{u_\mu(\bs{r})\}$. The positive-frequency part of the field can be decomposed as
\begin{align}
    \boldsymbol{E}^{(+)}(\bs{r},t) = \sum_\mu \mathcal{E}_\mu\, u_\mu(\bs{r})\, b_\mu\, e^{-i\omega_\mu t},
\end{align}
where $\{u_\mu(\bs{r})\}$ is an arbitrary orthonormal mode basis and $\mathcal{E}_\mu$ absorbs normalization factors. The time-averaged irradiance then reads
\begin{align}
    I(\bs{r}) \propto \sum_{\mu,\mu'} \mathcal{E}_\mu \mathcal{E}_{\mu'}^*\, u_\mu(\bs{r})\, u_{\mu'}^*(\bs{r})\, \braket{b_{\mu'}^\dagger b_\mu} \delta_{\omega_\mu,\omega_{\mu'}},
    \label{eq:irradiance_cross}
\end{align}
and when each mode has a different frequency, e.g. planar waves, cross terms vanish and Eq.~\eqref{eq:irradiance_cross} reduces to Eq.~\eqref{eq:EM_intensity_final}. When spatially distinct modes exist, the relative phase $u_\mu(\bs{r})\, u_{\mu'}^*(\bs{r})$ does not vanish and the cross terms contribute. In this work, we consider somewhat localized but progating modes and to avoid explicitly constructing the mode functions, we can approximate $u_\mu$ locally as planar modes that $u_\mu(\bs{r})\, u_{\mu'}^*(\bs{r})$ is the standard phase factor for spatally separated emitters which allows us to  recover the standard Airy reflectance formula.

\section{Thin-Film Formula from the Quantum Circuit}
\label{app:thin_film}

We derive Eq.~\eqref{eq:2L} directly by applying the collision circuit to the single-photon input state for $N_l=2$ layers. We write the state as $\ket{(n_{r_1} n_{r_2} \cdots)\; n_{s_1}\, n_m\, n_{s_2}\;(n_{t_1} n_{t_2} \cdots)}$, where the first parenthesis collects all incoming/reflected ray modes, the middle three entries are the surface mode of layer 1 ($s_1$), the intermediate traveling mode ($m$), and the surface mode of layer 2 ($s_2$), and the last parenthesis collects all transmitted modes. The initial state is
\begin{align}
    \ket{I} = \ket{(10\cdots)\;0\,0\,0\;(0\cdots)}.
\end{align}

Applying the first collision $U$ to the $(r_1, s_1, m)$ sector:
\begin{align}
    \ket{I} \;\xrightarrow{U}\; r\ket{(10\cdots)\;0\,0\,0\;(0\cdots)} + a\ket{(00\cdots)\;1\,0\,0\;(0\cdots)} + t\ket{(00\cdots)\;0\,1\,0\;(0\cdots)},
\end{align}
corresponding respectively to reflection into $r_1$, absorption into $s_1$, and transmission into the intermediate mode $m$.
As all modes propagate to the next collision, each component acquires a phase $e^{i\phi}$ where $\phi=-\omega d/\cos\theta$ is the free-evolution phase accumulated between the two layers:
\begin{align}
    re^{i\phi}\ket{(10\cdots)\;0\,0\,0\;(0\cdots)} + ae^{i\phi}\ket{(00\cdots)\;1\,0\,0\;(0\cdots)} + te^{i\phi}\ket{(00\cdots)\;0\,1\,0\;(0\cdots)}.
\end{align}

Then the second collision occurs on the $(m, s_2, t_1)$ sector, affecting only the transmitted component:
\begin{align}
    te^{i\phi}\ket{(00\cdots)\;0\,1\,0\;(0\cdots)} \;\xrightarrow{U}\; te^{i\phi}r\ket{(00\cdots)\;0\,1\,0\;(0\cdots)} + te^{i\phi}a\ket{(00\cdots)\;0\,0\,1\;(0\cdots)} + t^2e^{i\phi}\ket{(00\cdots)\;0\,0\,0\;(1\cdots)},
\end{align}
where the last term represents a photon fully transmitted through both layers, and $te^{i\phi}r\ket{(00\cdots)\;0\,1\,0\;(0\cdots)}$ is the photon reflected back from layer 2 into the intermediate mode, now heading toward layer 1.

After another phase pickup, $U$ acts simultaneously on two components: the absorbed state $ae^{i2\phi}\ket{(00\cdots)\;1\,0\,0\;(0\cdots)}$ and the back-reflected intermediate state $te^{i2\phi}r\ket{(00\cdots)\;0\,1\,0\;(0\cdots)}$. Both producing the first nontrivial round-trip contribution to the reflected sector:
\begin{align}
    te^{i2\phi}r\ket{(00\cdots)\;0\,1\,0\;(0\cdots)} +ae^{i2\phi}\ket{(00\cdots)\;1\,0\,0\;(0\cdots)}\;\xrightarrow{U}\; t^2re^{i2\phi}\ket{(01\cdots)\;0\,0\,0\;(0\cdots)} + a^2e^{i2\phi}\ket{(01\cdots)\;0\,0\,0\;(0\cdots)}+ \cdots,
\end{align}
This process continues iteratively: at each subsequent collision, new contributions from distinct optical paths accumulate in the reflected sector.

The final step is to evolve all the outgoing modes towards the detector located infinitely far away. Importantly, the modes are spatially separated so there is an additional phase pick up depending on the mode's index. In particular, each mode picks up a phase $e^{-i2\phi \sin^2 \theta}$ relative to the previous one. The final state is thus
\begin{align}
    r\ket{(10\cdots)\;0\,0\,0\;(0\cdots)} + (t^2re^{i\xi}+ a^2e^{i\xi})\ket{(01\cdots)\;0\,0\,0\;(0\cdots)}+ \cdots,
\end{align}
with $\xi = 2\phi (1-\sin^2 \theta)=-2 \omega d \cos\theta$. On which we can now evaluate the reflectance by summing over all reflected modes $k,j$:
\begin{align}
    R_\text{2L} = \braket{ \Bigl(\sum_k b^\dagger_k\Bigr)\Bigl(\sum_j b_j\Bigr)}
\end{align}

The cross terms between distinct paths reproduce Eq.~\eqref{eq:2L}, confirming that the quantum circuit recovers the standard thin-film Airy formula (with an additional absorption channel) for the single-photon input $\lvert 10\cdots\rangle$. To obtain the transmittance, it suffices to apply Eq.~\eqref{eq:EM_intensity_final} to the transmitted modes.

\begin{figure*}[t]
    \centering
    \includegraphics[width=\linewidth]{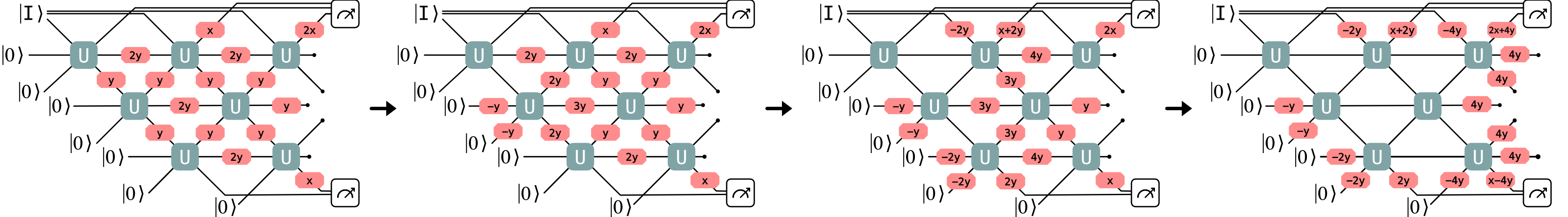}
    \caption{Left: the circuit used to compute the reflectance and transmittance of a multi-layer setup, where rectangular boxes denote phase operators $e^{-i\omega a\, b_j^\dagger b_j}$ with the value $a$ written inside. From left to right, commutation relations are applied to commute all phase operators past $U$, collecting them into the initial and final states only.}
    \label{fig:reduction}
\end{figure*}

The step-by-step protocol described above can be expressed as a quantum circuit, as illustrated by the first diagram of Fig.~\ref{fig:reduction}. Rectangular boxes denote the phase operator $e^{-i\omega a\, b_j^\dagger b_j}$, where $a$ is the value shown inside the box; concretely, $x = \phi$ and $y = -2\phi\sin^2\theta$ correspond to the in-layer and final phase pickups, respectively.

This reduction to the Airy summation is, in hindsight, expected: the circuit is nothing but a scattering formalism, and a single photon in a linear system must reproduce classical wave optics. The advantage of the quantum formulation becomes apparent when multiple rays are present, i.e. when the input state carries excitations across several modes, since the non-linear interaction terms $H_{\geq 4}$ then couple distinct photons in a way that has no classical analogue, generating the richer optical responses explored in the main text.

\section{Choice of Initial States}\label{app:initial_states}

A question raised in the main text was which initial state should we consider when describing multiple incident photons. There is no good answer since it depends on the type of light that is shining on the material. Normal light can be represented as a Gaussian state which, by definition, includes contributions from all photon-number sectors. Since we have capped the Fock space at a single excitation, it makes sense to discuss other possibilities for initial states.

The simplest possibility is to consider multiple independent photons in equal superposition, $\ket{I}=\ket{100 \cdots}+\ket{010 \cdots}+\ket{0010 \cdots}$. Since each term has a single particle, there will be no non-linear effect and each term will contribute the same $I_S$ to the (incident and reflected) irradiance so that the reflectance will not change. 

Another possibility is to consider the product state $\ket{I}=\prod_j b_j^\dagger\ket{0}$ which we did in the main text. In this state the two incident photons no longer have the maximum possible irradiance. As the light passes through the material, the reflected or transmitted irradiance can be larger than the original due to the presence of these cross terms. 
An interesting property of this state is that we can avoid evolving the state in between collisions, making our circuit shallower and easier to run on existing quantum hardware. To see that this is the case, let us analyze the circuit of Fig.~\ref{fig:reduction}, which explicitly accounts for the phase operators denoted by the rectangular boxes.

Recalling the symmetry constraints of $U$, we know that $U$ commutes with the total excitation number $\sum_j b^{\dagger}_jb_j$ where $j$ sums over the same modes as $U$ is acting on. This means it commutes with powers of $\exp(\sum_j b^{\dagger}_jb_j)$ and we can apply the commutation relation step by step from the start towards the end, as shown in Fig.~\ref{fig:reduction}. The result is that we only need to apply the phase operators to the initial and final state. But since the initial state is not in superposition, it picks up a global phase that does not change the final result (same for the material modes) and we only have to worry about the relative phases between the outgoing modes.
If we index the outgoing modes from left to right, the reflectance can be computed as
$\sum_{j,k}e^{i(k-j)\xi}\braket{ b^{\dagger}_jb_k}$ where the expectation value is evaluated on the final state of the circuit without the phase operators (which only needs to be evaluated once).

In the main work, we chose to keep $\ket{I}=\prod_j b_j^\dagger\ket{0}$ as part of the visual signature of our approach, but we could have equally explored other more exotic initial states that would have different visual outcomes.

\bibliographystyle{ACM-Reference-Format}
\bibliography{sample-base}
\end{document}